\documentclass[runningheads]{llncs}

\usepackage{microtype}

\usepackage{tikz}
\usepackage{amsmath}
\usepackage{tabu}
\usepackage{multirow}
\usepackage{booktabs}
\usepackage{color}
\usepackage[utf8]{inputenc}
\usepackage[binary-units]{siunitx}

\usepackage{graphicx}
\graphicspath{ {./img/} }

\newcommand\vTCollSep{\mathit{\$}}

\renewcommand{\log}{\lg}

\newcommand{\blk}{\textit{blk}}
\newcommand{\ptr}{\textit{ptr}}
\newcommand{\off}{\textit{off}}
\newcommand{\leafrank}{\textit{leaf-rank}}
\newcommand{\leafrankfield}{\textit{lrank}}
\newcommand{\startswithendleaf}{\textit{lbreaker}}
\newcommand{\minexcess}{\textit{min-excess}}
\newcommand{\minexcessfield}{\textit{mexcess}}
\newcommand{\leafselect}{\textit{leaf-select}}
\newcommand{\secondleafrank}{\textit{fb-lrank}}
\newcommand{\suffixstartswithendleaf}{\textit{fb-lbreaker}}
\newcommand{\mininfirstblock}{\textit{m-fb}}
\newcommand{\firstblockminexcess}{\textit{fb-mexcess}}

\newcommand{\rank}{\textit{rank}}
\newcommand{\select}{\textit{select}}
\newcommand{\excess}{\textit{excess}}
\newcommand{\bwdsearch}{\textit{bwd-search}}
\newcommand{\fwdsearch}{\textit{fwd-search}}
\newcommand{\secondrank}{\textit{fb-rank}}
\newcommand{\firstrank}{\textit{pfb-rank}}
\newcommand{\firstblockexcess}{\textit{fb-excess}}
\newcommand{\secondblockexcess}{\textit{sb-excess}}
\newcommand{\firstleafrank}{\textit{pfb-lrank}}
\newcommand{\secondblockminexcess}{\textit{sb-mexcess}}

\renewcommand{\log}{\lg}

\title{Faster Repetition-Aware Compressed Suffix Trees based on Block Trees%
\thanks{Partially funded by Fondecyt Grant 1-170048 and by Basal Funds FB0001, Conicyt. Chile.}}

\author{Manuel Cáceres\inst{1,2} \and Gonzalo Navarro\inst{1,2}}

\institute{CeBiB --- Center for Biotechnology and Bioengineering \and Department of Computer Science, University of Chile, Chile.\\
{\tt \{mcaceres, gnavarro\}@dcc.uchile.cl}}

\begin{document}

\maketitle

\begin{abstract}

Suffix trees are a fundamental data structure in stringology, but their space usage, though linear, is an important problem for its applications. We design and implement
 a new compressed suffix tree targeted to highly repetitive texts, such as large genomic collections of the same species. Our suffix tree builds on Block Trees, a recent Lempel-Ziv-bounded data structure that captures
 the repetitiveness of its input. We use Block Trees to compress the topology of the suffix tree, and augment the Block Tree nodes with data that speeds up suffix tree navigation.

 Our compressed suffix tree is slightly larger than previous repetition-aware suffix trees based on grammars, but outperforms them in time, often by orders of magnitude. The component that represents the tree topology achieves a speed comparable to that of general-purpose compressed trees, while using 2.3--10 times less space, and might be of interest in other scenarios.

\end{abstract}

\section{Introduction}
\label{sec:intro}

Suffix trees~\cite{suffixtreesweiner1973,suffixtreemccreight1976,suffixtreeukkonen1992} are one of the most appreciated data structures in
Stringology~\cite{suffixtreeapostolico1985} and in application areas like Bioinformatics \cite{suffixtreegusfield1997}, enabling efficient solutions to complex problems such as (approximate) pattern matching, pattern discovery, finding repeated substrings, computing matching statistics, computing maximal matches, and many others. In other collections, like natural language and software repositories, suffix trees are useful for plagiarism detection \cite{MFWJS05}, authorship attribution \cite{ZL06}, document retrieval \cite{HSTV14}, and others.

While their linear space complexity is regarded as acceptable in classical terms, their actual space usage brings serious problems in application areas. From an Information Theory standpoint, on a text of length $n$ over alphabet $[1,\sigma]$, classical suffix tree representations use $\Theta (n\log{n})$
bits, whereas the information contained in the text is, in the worst case, just
$n\log{\sigma}$ bits.
From a practical point of view, even carefully engineered implementations~\cite{suffixtreekurtz1999}
require at least $10$ bytes per symbol, which forces many applications to run the suffix tree on the (orders of magnitude slower) secondary memory.

Consider for example Bioinformatics, where various complex analyses require the use of sophisticated data structures, suffix trees being among the most important ones. DNA sequences range over $\sigma=4$ different nucleobases represented with $\log 4 = 2$ bits each, whereas the suffix tree uses at least $10$ bytes = $80$ bits
per base, that is, $4000\%$ of the text size. A human genome fits in approximately $715$ MB, whereas its suffix tree requires about $30$ GB. The space problem becomes daunting when we consider the DNA analysis of large groups of individuals; consider for example the 100,000-human-genomes project ({\tt www.genomicsengland.co.uk}).

One solution to the problem is to build suffix trees on secondary memory \cite{suffixtreeclark1996,stringbtreeferragina1999}. Most suffix tree algorithms, however, require traversing them across arbitrary access
paths, which makes secondary memory solutions many orders of magnitude slower than in main memory.
Another approach replaces the suffix trees with suffix arrays~\cite{suffixarraymanber1990}, which decreases space usage to $4$ bytes (32 bits) per character but loses some functionality like the suffix links, which are essential to solve various complex problems. This functionality can be recovered~\cite{suffixarrayabouelhoda2004} by raising the space to about $6$ bytes (48 bits) per character.

A promising line of research is the construction of compact representations of
suffix trees, named {\em Compressed Suffix Trees (CSTs)}, which simulate all the suffix tree functionality within space bounded not only by $O(n\log{\sigma})$ bits, but by the information content (or text entropy) of the sequence. An important theoretical achievement in this line was a CST using $O(n)$ bits on top of the text entropy that supports all the operations within an $O(\textrm{polylog}~n)$ time penalty factor~\cite{sadakane2007compressed}. A recent implementation~\cite{ohlebusch2010cst++} uses, on DNA, about 10 bits per base and supports the operations in a few microseconds. While even smaller CSTs have been proposed, reaching as little as 5 bits per base~\cite{russo2011fully}, their operation times raise until the milliseconds, thus becoming nearly as slow as a secondary-memory deployment.

Despite the progress done, further space reductions are desirable when facing large genome repositories. Fortunately
many of the largest text collections are highly repetitive; for example DNA sequences of two humans differ by less than $0.5\%$~\cite{similaritytishkoff2004}. This repetitiveness is not well captured
by statistical based compression methods~\cite{badentropykreft2013}, on which
most of the CSTs are based. Lempel-Ziv~\cite{lempel1976complexity} and
grammar~\cite{kieffer2000grammar} based compression
techniques, among others, do better in this scenario~\cite{repetitivenavarro2012}, but only recently we have seen CSTs building on them, both in theory~\cite{gagie2017optimal,belazzougui2017representing} and in practice \cite{abeliuk2013practical,navarro2016faster}.
The most successful CSTs in practice on repetitive collections are the grammar-compressed suffix trees (GCSTs), which on DNA use about 2 bits per base and support the operations in tens to hundreds of microseconds. 

GCSTs use grammar compression on the parentheses sequence that represents the suffix tree topology \cite{raman2013succinct}, which inherits the repetitiveness of the text collection. While Lempel-Ziv compression is stronger, it does not support easy access to the sequence. In this paper we explore an alternative to grammar compression called Block Trees~\cite{belazzougui2015queries,pereira2016statistical}, which offer similar approximation ratios to Lempel-Ziv compression, but promise faster access.

Our main algorithmic contribution is the BT-CT, a Block-Tree-based representation of tree topologies, which enriches Block Trees in order to support the required navigation operations. Although we are unable to prove useful upper bounds on the operation times, the BT-CT performs very well in practice: while using 0.3--1.5  bits per node in our repetitive suffix trees, it implements the navigation operations in a few microseconds, becoming very close to the performance of plain 2.8-bit-per-node representations that are blind to repetitiveness~\cite{navarro2014fully}. We use the BT-CT to represent suffix tree topologies in this paper, but it might also be useful in other scenarios, such as representing the topology of repetitive XML collections \cite{ACMNNSV13}.

As said, our new suffix tree, BT-CST, uses the BT-CT to represent the suffix tree topology.  Although larger than the GCST, it still requires about 3 bits per base in highly repetitive DNA collections. In exchange, it is faster than the GCST, often by an order of magnitude. This owes to the BT-CT directly, but also indirectly: Its faster navigation enables the binary search for the ``child by letter'' operation in suffix trees, which is by far the slowest one. While with the GCST a linear traversal of the children is advisable~\cite{navarro2016faster}, a binary search pays off in the BT-CST, making it faster especially on large alphabets.

\section{Preliminaries and Related Work}

A text  $T[1,n] = T[1]\ldots T[n]$ is a sequence of symbols over an alphabet $\Sigma =
[1,\sigma]$, terminated by a special symbol $\vTCollSep$ that is
lexicographically smaller than any symbol of $\Sigma$. A substring of $T$ is denoted $T[i,j]=T[i]\ldots T[j]$. A substring $T[i,j]$ is a prefix if $i=1$ and a suffix if $j=n$.

The {\em suffix tree}~\cite{suffixtreesweiner1973,suffixtreemccreight1976,suffixtreeukkonen1992} of a text $T$ is a trie of its suffixes
in which unary paths are collapsed into a single edge. The tree then has less than $2n$ nodes. The suffix tree supports a set of operations (see Table~\ref{tab:st}) that suffices to solve a large number of problems in Stringology~\cite{suffixtreeapostolico1985} and Bioinformatics~\cite{suffixtreegusfield1997}.
\begin{table}
\centering
\begin{tabular}{ l l }
  \textbf{Operation} & \textbf{Description}  \\
  \hline
  root() & The root of the suffix tree  \\
  is-leaf($v$) & True if $v$ is a leaf node \\
  first-child($v$) & The first child of $v$ in lexicographical order  \\
  tree-depth($v$) & The number of edges from root() to $v$\\
  next-sibling($v$) & The next sibling of $v$ in lexicographical order \\
  previous-sibling($v$) & The previous sibling of $v$ in lexicographical order \\
  parent($v$) & The parent of $v$ \\
  is-ancestor($v$,$u$) & True if $v$ is ancestor of $u$\\
  level-ancestor($v$,$d$) & The ancestor of $v$ at tree depth $d$\\
  lca($v$,$u$) & The lowest common ancestor between $v$ and $u$\\
  letter($v$, $i$) & str($v$)[$i$]\\
  string-depth($v$) & $|$str($v$)$|$ \\
  suffix-link($v$) & The node $u$ s.t. str($u$) = str($v$)[2,string-depth($v$)]\\
  string-ancestor($v$,$d$) & The highest ancestor $u$ of $v$ s.t. string-depth($u$) $\ge d$\\
  child($v$,$c$) & The child $u$ of $v$ s.t. str($u$)[string-depth($v$)+1] = $c$\\ \ \\
\end{tabular}
\caption{List of typical operations implemented by suffix trees; str($v$) represents the concatenation of the strings in the root-to-$v$ path.}
\label{tab:st}
\end{table}

The {\em suffix array}~\cite{suffixarraymanber1990} $A[1,n]$ of a text $T[1,n]$ is a
permutation of $[1,n]$ such that $A[i]$ is the starting position of the $i$th suffix in increasing lexicographical
order. Note that the leaves descending from a suffix tree node span a range of suffixes in $A$.

The function $lcp(X,Y)$ is the length of the longest common prefix (lcp) of strings $X$ and $Y$. The {\em LCP array}~\cite{suffixarraymanber1990}, $LCP[1,n]$, is defined as $LCP[1]=0$ and $LCP[i]=lcp(T[A[i-1],n],T[A[i],n])$ for all $i >  1$, that is, it stores the lengths of the longest common prefixes between lexicographically consecutive suffixes of $T[1,n]$.

\subsection{Succinct tree representations}
\label{sec:bp}

A balanced parentheses (BP) representation (there are others~\cite{raman2013succinct}) of the topology of an ordinal tree
$\mathcal{T}$ of $t$ nodes is a binary sequence (or bitvector) $P[1,2t]$ built as follows: we traverse $\mathcal{T}$
in preorder, writing an opening parenthesis (a bit 1) when we first arrive at a node, and a
closing one (a bit 0) when we leave its subtree. For example, a leaf looks like ``10''. The following primitives can be defined on $P$:

\begin{itemize}
\item $access(i) = P[i]$
\item $\rank_{0|1}(i) = |\left\{1\le j \le i; P[j] = 0|1\right\}|$
\item $\excess(i) = \rank_{1}(i)-\rank_{0}(i)$
\item $\select_{0|1}(i) = \min (\{j; \rank_{0|1}(j) = i\} \cup \{\infty\})$
\item $\leafrank(i) = \rank_{10}(i) = |\left\{1\le j \le i-1; P[j] = 1 \land P[j+1] = 0\right\}|$
\item $\leafselect(i) = \select_{10}(i) = \min (\{j; \leafrank(j+1) = i\} \cup \{\infty\})$
\item $\fwdsearch(i,d) = \min (\{j>i; \excess(j) = \excess(i) + d)\} \cup \{\infty\})$
\item $\bwdsearch(i,d) = \max (\{j<i; \excess(j) = \excess(i) + d)\} \cup \{-\infty\})$
\item $\minexcess(i,j) = \min (\{\excess(k)-\excess(i-1); i \le k\le j\} \cup \{\infty\})$
\end{itemize}

These primitives suffice to implement a large number of tree navigation operations, and can all be supported in constant time using $o(t)$ bits on top of $P$~\cite{navarro2014fully}. These include the operations needed by suffix trees. For example, interpreting nodes as the position of their opening parenthesis in $P$, it holds that $parent(v) = \bwdsearch(i, -2)+1$, $\textit{next-sibling}(v) = \fwdsearch(v,-1)+1$
and the lowest common ancestor of two nodes $v \le u$ is $lca(v, u) = parent(\fwdsearch(v-1, \minexcess(v,u))+1)$.

\subsection{Compressed Suffix Arrays}

A milestone in the area was the emergence of Compressed Suffix Arrays (CSAs)~\cite{csanavarro2007},
which using space proportional to that of the compressed sequence managed to answer access queries
to the original suffix array and its inverse (i.e., return any $A[i]$ and $A^{-1}[j]$), to the indexed sequence (i.e., return any $T[i..j]$), and access to a novel array,
$\Psi[i] = A^{-1}[(A[i]\!\!\mod\ n)+1]$, which lets us move from a text suffix $T[j,n]$ to the next one, $T[j+1,n]$, yet indexing the suffixes by their lexicographic rank, $A^{-1}[j]$. This function plays an important
role in the design of CSTs, as we see next.

\subsection{Compressed Suffix Trees}

Sadakane~\cite{sadakane2007compressed} designed the first CST, on top of a CSA, using $|\mathrm{CSA}|+O(n)$ bits and solving all the suffix tree operations in time $O(\textrm{polylog}~n)$.
He makes up a CST from three components: a CSA, for which he uses his
own proposal~\cite{sadakane2003new}; a BP  representation of the suffix tree topology, using at most $4n+o(n)$ bits; and a compressed
representation of $LCP$, which is a bitvector $H[1,2n]$ encoding the array $PLCP[i] = LCP[A^{-1}[i]]$ (i.e., the LCP array in text order). A recent implementation~\cite{ohlebusch2010cst++} of this index requires about $10$
bits per character and takes a few microseconds per operation.

Russo et~al.~\cite{russo2011fully} managed to use just $o(n)$ bits on top of the CSA, by storing only a sample of the suffix tree nodes. An implementation of this index~\cite{russo2011fully}  uses as little as 5 bits per character, but the operations take milliseconds, as slow as running in secondary storage.

Yet another approach~\cite{fischer2009faster} also obtains $o(n)$ on top of a CSA by getting rid of the tree topology and expressing the tree operations on the corresponding suffix array intervals. The operations now use primitives on the LCP array: find the previous/next smaller value (psv/nsv) and find minima in ranges (rmq). They also noted that bitvector $H$ contains $2r$ runs, where $r$
is the number of runs of consecutive increasing values in $\Psi$, and used this fact to run-length compress
$H$. Abeliuk et~al.~\cite{abeliuk2013practical} designed a practical version of this idea, obtaining
about $8$ bits per character and getting a time performance of hundreds of microseconds per operation,
an interesting tradeoff between the other two options.

Engineered adaptations of these three ideas were implemented in the SDSL library~\cite{gogcompressed}, and are named \texttt{cst\_sada}, \texttt{cst\_fully},
and \texttt{cst\_sct3}, respectively. We will use and adapt them in our experimental comparison.

\subsection{Repetition-aware Compressed Suffix Trees}\label{rel_work}

Abeliuk et$.$ al~\cite{abeliuk2013practical} also presented the first CST for
repetitive collections. They built on the third approach above~\cite{fischer2009faster}, so they do not represent the tree topology. They use the RLCSA~\cite{rlcsamakinen2010storage}, a
repetition-aware CSA with size proportional to $r$, which is very low on repetitive texts. They use grammar compression on the {\em differential} LCP array, $DLCP[i]=LCP[i]-LCP[i-1]$. The nodes of the parsing tree (obtained with Re-Pair~\cite{repairlarsson1999}) are enriched with further data to support the operations psv/nsv and rmq. To speed up simple LCP accesses, the bitvector $H$ is also stored, whose size is also proportional to $r$. Their index uses 1--2 bits per character on repetitive collections.
It is rather slow, however, operating within (many) milliseconds.

Navarro and Ord\'o\~nez~\cite{navarro2016faster} include again the tree topology. Since text repetitiveness induces isomorphic subtrees in the suffix tree, they grammar-compressed the BP representation. The nonterminals are enriched to support the tree navigation operations enumerated in Section~\ref{sec:bp}. Since they do not need
psv/nsv/rmq operations on LCP, they just use the bitvector $H$, which has a few runs and thus is very small. Their index uses slightly more space, closer to 2 bits per character, but it is up to 3 orders of magnitude faster than that of Abeliuk et~al.~\cite{abeliuk2013practical}: their structure operates in tens to hundreds of microseconds per operation,
getting closer to the times of general-purpose CSTs.

It is worth mentioning the recent work by Farruggia et~al.~\cite{farruggia2015relative},
who builds on Relative Lempel-Ziv~\cite{rlzkuruppu2010} to compress
the suffix trees of the individual sequences (instead of that of the whole collection). They showed to be time- and space-competitive against the CSTs mentioned, but their structure offers a different functionality (useful for other problems). Some recent theoretical work includes Gagie et~al.'s~\cite{gagie2017optimal} $O(r\log(n/r)\log n)$-bits CST using Run-Length Context-Free Grammars~\cite{nishimoto2016fully} on the LCP and
the suffix arrays, and Belazzougui and Cunial's~\cite{belazzougui2017representing} CST based on the CDAWG~\cite{belazzougui2015composite} (a minimized automaton that recognizes all the substrings of $T$). Both support most operations in time $O(\log n)$.

\section{Block Trees}
\label{sec:bt}

A Block Tree~\cite{belazzougui2015queries} is a full $r$-ary tree that represents a (repetitive) sequence $P[1,p]$ in compressed space while offering access and other operations in logarithmic time. The nodes at depth $d$ (the root being depth 0) represent blocks of $P$ of length $b=|P|/r^d$, where we pad $P$ to ensure these numbers are integers. Such a node $v$, representing some block $v.\blk = P[i,i+b-1]$, can be of three types:

\begin{description}
\item[LeafBlock:] If $b \le mll$, where $mll$ is a parameter, then $v$ is a leaf of the Block Tree, and it stores the string $v.blk$ explicitly.
\item[BackBlock:] Otherwise, if $P[i-b,i+b-1]$ and $P[i,i+2b-1]$ are not their leftmost occurrences in $P$, then the block is replaced by its leftmost occurrence in $P$: node $v$ stores a pointer $v.\ptr=u$ to the node $u$ such that the first occurrence of $v.\blk$ starts inside $u.\blk = P[j,j+b-1]$, more precisely it occurs in $P[j+o,j+o+b-1]$. This offset inside $u.\blk$ is stored at $v.\off=o$. Node $v$ is not considered at deeper levels.
\item[InternalBlock:] Otherwise, the block is split into $r$ equal parts, handled in the next level by the children of $v$. The node $v$ then stores a pointer to its children.
\end{description}

The Block Tree can return any $P[i]$ by starting at position $i$ in the root block. Recursively, the position $i$ is translated in constant time into an offset inside a child node (for InternalBlocks), or inside a leftward node in the same level (for BackBlocks, at most once per level). At leaves, the symbol is stored explicitly.
Setting $mll = \log_\sigma{p}$ and removing the first $\log_{r}z$
levels, where $z$ is the size of the Lempel-Ziv parse of $P$, the following properties hold:
%
(1) The height of the Block Tree is $O\left(\log_r{\frac{p\log{\sigma}}{z\log{p}}}\right)$.
(2) The Block Tree uses $O\left(zr\log{p} \log_r{\frac{p\log{\sigma}}{z\log{p}}}\right)$ bits of space.
(3) The target nodes of BackBlocks are always InternalBlocks, and thus access takes
$O\left(\log_r{\frac{p\log{\sigma}}{z\log{p}}}\right)$ time.

If we augment the nodes of the Block Tree with rank information for the $\sigma$ symbols of the alphabet, the Block Tree now uses $O\left(\sigma zr\log{p} \log_r{\frac{p\log{\sigma}}{z\log{p}}}\right)$
bits and can answer rank and select queries on $P$ in time $O\left(r\log_r{\frac{p\log{\sigma}}{z\log{p}}}\right)$.
Specifically, for every $c \in [1,\sigma]$, we store in every node $v$ the number $v.c$ of $c$s in $v.\blk$. Further, every BackBlock node $v$ pointing to $u$ stores the number of $c$s in $u.\blk[1,v.\off-1]$.

Our new repetition-aware CST will represent the BP topology
with a Block Tree. Since in this case $\sigma = 2$ and $n \le p \le 4n$, the Block Tree will use
$O\left(zr\log{n} \log_r{\frac{n}{z\log{n}}}\right)$ bits of space and answer
$access(i)$, $\rank_{0|1}(i)$, $\excess(i)$ and $\select_{0|1}(i)$ in time $O\left(r\log_r{\frac{n}{z\log{n}}}\right)$.
In the next section we show how to solve the missing operations ($\leafrank(i)$,
 $\leafselect(i)$, $\fwdsearch(i,d)$, $\bwdsearch(i,d)$ and $\minexcess(i,j)$) by storing further data on the Block Tree nodes, yet preserving the $O\left(zr\log{n} \log_r{\frac{n}{z\log{n}}}\right)$-bits space.

\section{Our Repetition-Aware Compressed Suffix Tree}

Following the scheme of Sadakane~\cite{sadakane2007compressed} we propose a three-component structure
to implement a new CST tailored to highly repetitive inputs.
We use the RLCSA~\cite{rlcsamakinen2010storage} as our CSA. For the LCP, we use the compressed version
of the bitvector $H$~\cite{fischer2009faster}. For the topology, we use BP and represent the sequence with a Block Tree, adding new fields to the Block Tree nodes to efficiently answer all the queries we need (Section~\ref{sec:bp}). We call this representation Block Tree CST (BT-CST).
Section~\ref{sec:bt-ct} describes BT-CT, our extension to Block Trees, and Section~\ref{sec:child} our improved operation child($v,a$) for the BT-CST.

\subsection{Block Tree Compressed Topology (BT-CT)} \label{sec:bt-ct}

We describe our main data structure, {\em Block Tree Compressed Topology (BT-CT)}, which compresses a parentheses sequence and supports navigation on it.

\subsubsection{Block Tree augmentation}

\paragraph{\textbf{Stored fields.} \rm We augment the nodes of the Block Tree with the following fields:}
\begin{itemize}
\item For every node $v$ that represents the block $v.\blk=P[i,i+b-1]$:
\begin{itemize}
\item $\rank_1$, the number of 1s in $v.\blk$, that is, $\rank_1(i+b-1)-\rank_1(i-1)$ in $P$.
\item $\leafrankfield$ (leaf rank), the number of $10$s (i.e., leaves in BP) that finish inside $v.\blk$, that is, $ \leafrank(i+b-1) - \leafrank(i-1)$ in $P$.
\item $\startswithendleaf$ (leaf breaker), a bit telling whether the first symbol of $v.\blk$ is a $0$ and the preceding symbol in $P$ is a $1$, that is, whether $P[i-1,i] = 10$.
\item $\minexcessfield$, the minimum excess reached in $v.\blk$, that is, $\minexcess(i,i+b-1)$ in $P$.
\end{itemize}
\item For every BackBlock node $v$ that represents $v.\blk = P[i,i+b-1]$ and points to its first occurrence $O=P[j+o,j+o+b-1]$ inside $u.\blk = P[j,j+b-1]$ with offset $v.\off=o$:
\begin{itemize}
\item $\secondrank_1$, the number of 1s in the prefix of $O$ contained in $u.\blk$ ($O \cap u.\blk$, the first block spanned by $O$), that is,
$\rank_1(j+b-1)-\rank_1(j+o-1)$ in $P$.
\item \secondleafrank, the number of $10$s that finish in $O \cap u.\blk$, that is,
$\leafrank(j+b-1)-\leafrank(j+o-1)$ in $P$.
\item \suffixstartswithendleaf, a bit telling whether the first symbol of $O$ is a $0$ and
the preceding symbol is a $1$, that is, whether
$P[j+o-1,j+o] = 10$.
\item \firstblockminexcess, the minimum excess
reached in $O \cap u.\blk$, that is, $\minexcess(j+o,j+b-1)$.
\item \mininfirstblock, a bit telling whether the minimum excess of $u.\blk$ is reached in $O \cap u.\blk$, that is, whether
$\minexcess(i,i+b-1) = \minexcess(j+o,j+b-1)$.
\end{itemize}
\end{itemize}

\paragraph{\textbf{Fields computed on the fly.} \rm
In the description of the operations we will use other fields that are computed in constant time from those we already store:}

\begin{itemize}
\item For every node $v$ that represents $v.\blk = P[i,i+b-1]$
\begin{itemize}
\item $\rank_0$, the number of $0$s in $v.\blk$, that is, $b-v.\rank_1$.
\item $\excess$, the excess of 1s over 0s in $v.\blk$, that is, $v.\rank_1 - v.\rank_0 = 2\cdot v.\rank_1 - b$.
\end{itemize}
\item For every BackBlock node $v$ that represents $v.\blk = P[i,i+b-1]$ and points to its first occurrence $O=P[j+o,j+o+b-1]$ inside $u.\blk = P[j,j+b-1]$ with offset $v.\off=o$:
\begin{itemize}
\item $\secondrank_0$, the number of 0s in $O \cap v.\blk$, that is, $(b-o)-v.\secondrank_1$.
\item $\firstrank_{0|1}$, the number of 0s$|$1s  in the prefix of $u.\blk$ that precedes $O$ ($u.\blk - O$), that is, $u.\rank_{0|1} - v.\secondrank_{0|1}$.
\item $\firstblockexcess$, the excess in $O \cap u.\blk$, that is, $v.\secondrank_1 - v.\secondrank_0$.
\item $\secondblockexcess$, the excess in $O - u.\blk$ (2nd block spanned by $O$), that is, $v.\excess - v.\firstblockexcess$.
\item $\firstleafrank$, the number of $10$s that finish in $u.\blk - O$, that is, $u.\leafrankfield - v.\secondleafrank$.
\item $\secondblockminexcess$, the minimum excess in $O-u.\blk$, that is, $\minexcess(j+b,j+b+o-1)$ in $P$. We store either $v.\firstblockminexcess$ or $v.\secondblockminexcess$, the one that differs from $v.\minexcessfield$. To deduce the non-stored field we use
$\minexcessfield$, $\firstblockexcess$ and $\mininfirstblock$.
\end{itemize}
\end{itemize}

\subsubsection{Complex operations}

Apart from the basic operations solved in the original Block Tree we need, as described in Section~\ref{sec:bp}, more sophisticated ones to support navigation in the parentheses sequence.

\paragraph{\textbf{$\leafrank(i)$ and $\leafselect(i)$.} \rm
The implementations of these operations are analogous to those for $\rank_c(i)$ and $\select_c(i)$
respectively, in the base Block Tree. The only two differences are that in LeafBlocks we consider
the $\startswithendleaf$  field to check whether the block starts with a leaf, and in BackBlocks
we consider fields $\startswithendleaf$ and $\suffixstartswithendleaf$ to check
whether we have to add or remove one leaf when moving to a leftward node.
Like $\rank_c(i)$ and $\select_c(i)$, our operations work $O(1)$ per level, and then have their same time complexity, given in Section~\ref{sec:bt}.}

\paragraph{\textbf{$\fwdsearch(i,d)$ and $\bwdsearch(i,d)$.} \rm
We only show how to solve $\fwdsearch(i,d)$ with $d<0$; the other cases are similar (some combinations not needed for our CST require further fields). Thus we aim to find the smallest position $j>i$ where the excess of $P[i+1..j]$ is $d$. \\}

We describe our solution as a recursive procedure $\fwdsearch(i, j)$ with two global
variables: $d$ from the input, and $e$. Variables $i$ and $j$ are the limits of the search for the currently processed node, and $e$ is the accumulated excess of the part of the
range that has already been processed. The procedure is initially called
at the Block Tree root with $\fwdsearch(i,n)$ and with $e=0$. If at some point $e$ reaches $d$,
we have found
the answer to the search. The general idea is to traverse the range of the current node $v$ left to right, using the fields $v.\minexcessfield$, $v.\firstblockminexcess$ and $v.\secondblockminexcess$ to speed up the procedure:
\begin{itemize}
\item If the search range spans the entire block $v.\blk$ (i.e., $j-i = b$) and the answer
is not reached inside $v$ (i.e., $e+v.\minexcessfield>d$), then we increase $e$ by $v.\excess$ and return $\infty$.
\item If $v$ is a LeafBlock we scan $v.\blk$ bitwise, increasing $e$ for each 1 and decreasing $e$ for each 0. If $e$ reaches $d$ at some index $k$, we return
$k$; otherwise we return $\infty$.
\item If $v$ is an InternalBlock, we identify the $k$-th child of $v$, which contains position $i+1$, and the $m$-th, which contains position $j$ (it could be that $k=m$). We then call $\fwdsearch$ recursively on the $k$-th to the $m$-th children, intersecting the query range with the extent of each child (the search range will completely cover the children after the $k$-th and before the $m$-th). As soon as any of these calls returns a non-$\infty$ value, we adjust (i.e., shift) and return it. If all of them return $\infty$, we also return $\infty$.
\item If $v$ is a BackBlock we must translate the query to the original block $O$, which starts at offset $v.\off$ in $u.\blk$, where $u=v.\ptr$. We first check whether the query covers the prefix of $v.\blk$ contained in $u.\blk$, $O \cap u.\blk$ (i.e., if $i = 0$ and $j \ge b-v.\off$). If so, we check whether we can skip $O \cap u.\blk$, namely if $e+v.\firstblockminexcess > d$. If we can skip it, we just update $e$ to $e+v.\firstblockexcess$, otherwise we call $\fwdsearch$ recursively on the intersection of $u.\blk$ and the translated query range. If the answer is not $\infty$, we adjust and return it. Otherwise, we turn our attention to the node $u'$ next to $u$. Again, we check whether the query covers the suffix of $v.\blk$ contained in $u'.\blk$, $O - u.\blk$ (i.e., $j = b$ and $i \le b-v.\off$). If so, we check whether we can skip $O - u.\blk$, namely if $e + v.\secondblockminexcess > d$. If we can skip it, we just update $e$ to $e+v.\secondblockexcess$, otherwise we call $\fwdsearch$ recursively on the intersection of $u'.\blk$ and the translated query range. If the answer is not $\infty$, we adjust and return it. Otherwise, we return $\infty$.
\end{itemize}

\paragraph{\textbf{$\minexcess(i,j)$.} \rm
This operation seeks the minimum excess in $P[i..j]$. We will also start at the root with the global variable $e$ set to zero. A local variable $m$ will keep track of the minimum excess seen in the current node, and will be initialized at $m=1$ in each recursive call. The idea is the same as for $\fwdsearch$: traverse the node left to right and use the fields $v.\minexcessfield$, $v.\firstblockminexcess$ and $v.\secondblockminexcess$ to speed up the traversal.}

\begin{itemize}
\item If the query covers the entire block $v.\blk$ (i.e., $j-i+1 = b$), we increase $e$ by $v.\excess$ and return $v.\minexcessfield$.
\item If $v$ is a LeafBlock we record the initial excess in $e'=e$ and scan $v.\blk$ bitwise, updating $e$ for each bit read as in operation $\fwdsearch$. Every time we have $e-e'<m$, we update $m = e-e'$. At the end of the scan we return $m$.

\item If $v$ is an InternalBlock, we identify the $k$-th child of $v$, which contains position $i$, and the $m$-th, which contains position $j$ (it could be that $k=m$). We then call $\minexcess$ recursively on the $k$-th to the $m$-th children, intersecting the query range with the extent of each child (the search range will completely cover the children after the $k$-th and before the $m$-th, so these will take constant time). We return the minimum between all their answers (composed with their correspondent prefix excesses).

\item If $v$ is a BackBlock we translate the query to the original block $O$, which starts at offset $v.\off$ in $u.\blk$, where $u=v.\ptr$. We first check whether the query covers the prefix of $v.\blk$  contained in  $u.\blk$, $O \cap u.\blk$ (i.e., if $i = 1$ and $j \ge b-v.\off-1$). If so, we simply set $m = v.\firstblockminexcess$ and update $e$ to $e+v.\firstblockexcess$. Otherwise we call $\minexcess$ recursively on the intersection of $u.\blk$ and the translated query range, and record its answer in $m$. We now consider the block $u'$ next to $u$ and again check whether the query covers the suffix of $v.\blk$ contained in $u'.\blk$, $O-u.\blk$ (i.e., if $j = b$ and $i \le b-v.\off+1$). If so, we just set $m = \min(m, v.\firstblockexcess + v.\secondblockminexcess)$ and update $e$ to $e+v.\secondblockexcess$. Otherwise, we call $\minexcess$ on the intersection of $u'.\blk$ and the translated query range, record its answer in $m'$, and set $m = \min(m,v.\firstblockexcess+m')$. Finally, we return $m$.
\end{itemize}

\smallskip
Note that, although we look for various opportunities of using the precomputed data to skip parts of the query range, the operations $\fwdsearch$, $\bwdsearch$, and $\minexcess$ are not guaranteed to work proportionally to the height of the Block Tree. The instances we built that break this time complexity, however, are unlikely to occur. Our experiments will show that the algorithms perform well in practice.

\subsection{Operation child} \label{sec:child}

The fast operations enabled by our BT-CT structure give space for an improved algorithm to solve operation child($v,a$). Previous CSTs first compute $d=\textrm{string-depth}(v)$ and then linearly traverse the children of $v$ from $u=\textrm{first-child}(v)$ with operation next-sibling, checking for each child $u$ whether $\textrm{letter}(u,d+1)=a$, and stopping as soon as we find or exceed $a$. Since computing letter is significantly more expensive than our next-sibling, we consider the variant of first identifying all the children $u$ of $v$, and then binary searching them for $a$, using letter. We then perform $O(\sigma)$ operations next-sibling, but only $O(\log\sigma)$ operations letter.

\section{Experiments and Results}

We measured the time/space performance of our new BT-CST and compared it with the state of the art.
Our code and testbed will be available at \\{\tt https://github.com/elarielcl/BT-CST}.

\subsection{Experimental setup}

\paragraph{\textbf{Compared CSTs.} \rm We compare the following CST implementations. }
\begin{description}
    \item[BT-CST.] Our new Compressed Suffix Tree with the described components. For the BT-CT component we vary $r \in \{2,4,8\}$ and $mll \in \{4, 8, 16, 32, 64, 128, 256\}$.
    \item[GCST.] The Grammar-based Compressed Suffix Tree~\cite{navarro2016faster}. We vary parameters {\em rule-sampling} and {\em C-sampling} as they suggest.
    \item[CST\_SADA ,CST\_SCT3, CST\_FULLY.] Adaptation and improvements from the SDSL library\footnote{Succinct data structures library (SDSL), \url{https://github.com/simongog/sdsl-lite}} on the indexes of Sadakane~\cite{sadakane2007compressed}, Fischer et~al.~\cite{fischer2009faster} and Russo et~al.~\cite{russo2011fully}, respectively. \textsf{CST\_SADA} maximizes speed using Sadakane's CSA~\cite{sadakane2003new} and a non-compressed version of bitvector $H$. \textsf{CST\_SCT3} uses instead a Huffman-shaped wavelet tree of the BWT as the suffix array, and a compressed representation \cite{raman2007succinct} for bitvector $H$ and those of the wavelet tree. This bitvector representation exploits the runs and makes the space sensitive to repetitiveness, but it is slower. \textsf{CST\_FULLY} uses the same BWT representation. For all these suffix arrays we set $\textit{sa-sampling} = 32$ and $\textit{isa-sampling} = 64$.
    \item[CST\_SADA\_RLCSA, CST\_SCT3\_RLCSA.] Same as the preceding implementations but (further) adapted to repetitive collections: We replace the suffix array by the RLCSA~\cite{rlcsamakinen2010storage} and use a run-length-compressed representation of bitvector $H$~\cite{fischer2009faster}.
\end{description}

For the CSTs using the RLCSA, we fix their parameters to $32$ for the sampling of $\Psi$ and $128$ for the text sampling. We only show the Pareto-optimal results of each structure. Note that we do not include the CST of Abeliuk et al.~\cite{abeliuk2013practical} in the comparison because it was already outperformed by several orders of magnitude by \textsf{GCST}~\cite{navarro2016faster}.

\paragraph{\textbf{Text collection and queries.} \rm
Our input sequences come from the Repetitive Corpus of \textit{Pizza\&Chili} (\url{http://pizzachili.dcc.uchile.cl/repcorpus}). We selected \texttt{einstein}, containing all the versions (up to January 12, 2010) of the German Wikipedia Article of \textit{Albert Einstein} (89MB, compressible by p7zip to 0.11\%); \texttt{influenza}, a collection of 78,041 H. influenzae genomes (148MB, compressible by p7zip to 1.69\%); and \texttt{kernel}, a set of 36 versions of the Linux Kernel (247MB, compressible by p7zip to 2.56\%). \\}

Data points are the average of 100,000 random queries, following the scheme
used in previous repetition-aware CSTs~\cite{abeliuk2013practical,navarro2016faster} to choose the nodes on which the operations are called.

\paragraph{\textbf{Computer.} \rm
The experiments ran on an isolated Intel(R) Xeon(R) CPU E5-2407 @ 2.40GHz
with 256GB of RAM and 10MB of L3 cache. The operating system is GNU/Linux, Debian 2, with kernel 4.9.0-8-amd64. The implementations use a single thread and all of
them are coded in C++. The compiler is gcc version 4.6.3, with
-O9 optimization flag set (except \textsf{CST\_SADA}, \textsf{CST\_SCT3} and \textsf{CST\_FULLY}, which use their own set of optimization flags).}

\paragraph{\textbf{Operations.} \rm
We implemented all the suffix tree operations of Table~\ref{tab:st}. From those, for lack of space, we present the performance comparison with other CSTs on five important operations: next-sibling, parent, child, suffix-link, and lca.}

To test our suffix tree in more complex scenarios we implemented the suffix-tree-based algorithm to solve the ``maximal substrings'' problem~\cite{navarro2016faster} on all of the above implementations except for \textsf{CST\_FULLY} (because of its poor time performance). We use their same setup~\cite{navarro2016faster}, that is,  \texttt{influenza} from \textit{Pizza\&Chili} as our larger sequence and a substring of size $m$ ($m=3000$ and $m=2$MB) of another \texttt{influenza} sequence taken from \url{https://ftp.ncbi.nih.gov/genomes/INFLUENZA}. \textsf{BT-CST} uses $r=2$ and $mll=128$ and \textsf{GCST} uses {\em rule-sampling} $=1$ and {\em C-sampling} $=2^{10}$. The tradeoffs refer to {\em sa-sampling} $\in \{ 64,128,256\}$ for the RLCSAs.

\subsection{Results and discussion}

\begin{figure}[p]
    \centering
    \vspace*{-9mm}

    \hspace*{-2mm}
    \includegraphics[scale=0.33]{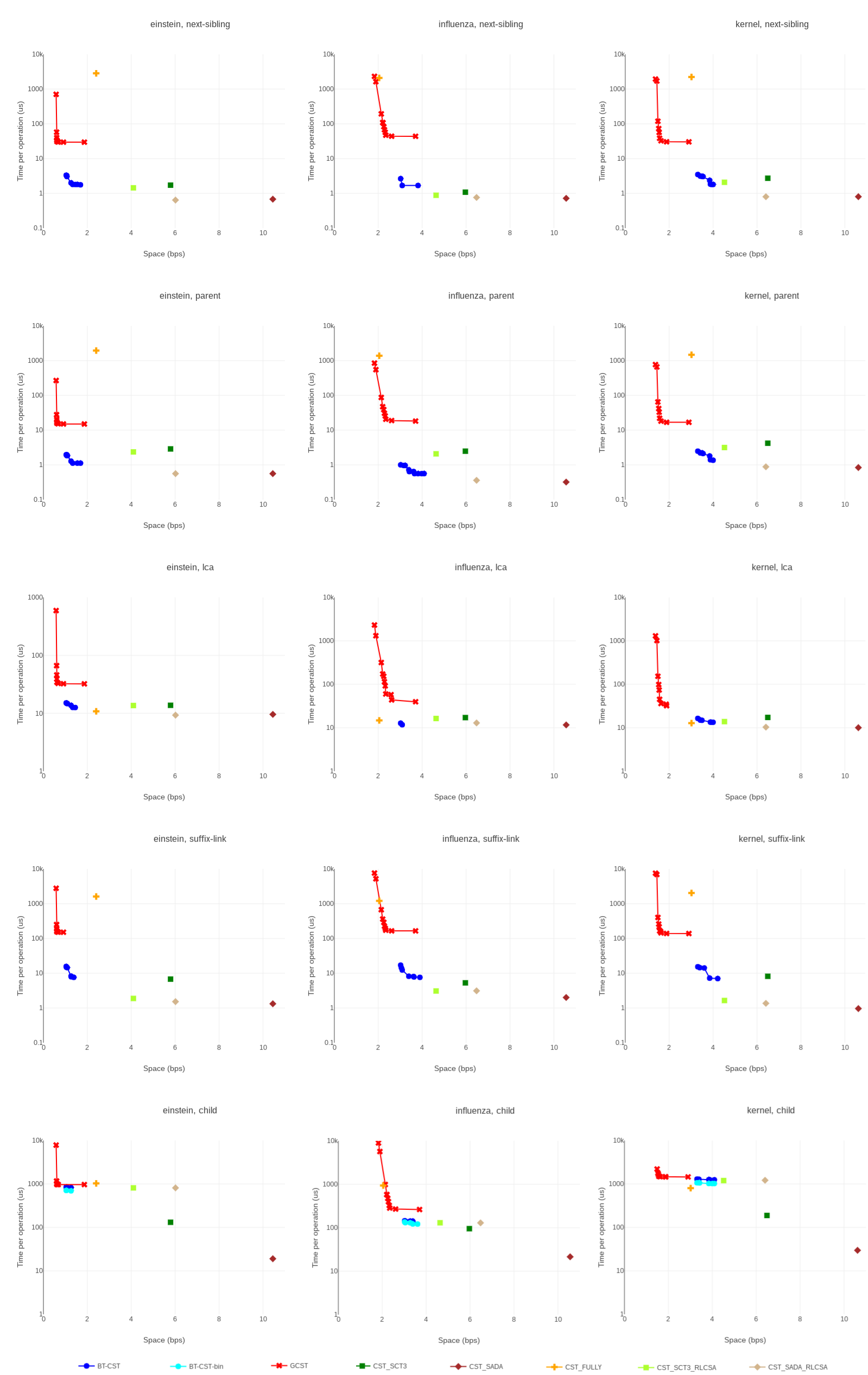}
    \caption{Performance of CSTs for operations next-sibling, parent, lca, suffix-link, and child. The y-axis is time in microseconds in log-scale. \textsf{BT-CST-bin} is \textsf{BT-CST} with binary search for child.}
    \label{operations}
\end{figure}

Figure~\ref{operations} shows the space and time for all the indexes and all the operations. The smallest structure is \textsf{GCST}, which takes as little as 0.5--2 bits per symbol (bps). The next smallest indexes are \textsf{BT-CST}, using 1--3 bps, and \textsf{CST\_FULLY}, using 2.0--2.5 bps. The compressed indexes not designed for repetitive collections use 4--7 bps if combined with a RLCSA, and 6--10.5 bps in their original versions (though we also adapted the bitvectors of \textsf{CST\_SCT3}).

From the \textsf{BT-CST} space, component $H$ takes just 2\%--9\%, the RLCSA takes 23\%--47\%, and the rest is the BT-CT (using a sweetpoint configuration). This component takes 0.30 bits per node (bpn) on \texttt{einstein}, 1.06 bpn on \texttt{influenza}, and 1.50 bpn on \texttt{kernel}. The grammar-compressed topology of \textsf{GCST} takes, respectively, 0.05, 0.81, and 0.39 bpn.

In operations next-sibling and parent, which rely most heavily on the suffix tree topology, our BT-CT component building on Block Trees makes \textsf{BT-CST} excel in time: The operations take nearly one microsecond ($\mu$sec), at least 10 times less than the grammar-based topology representation of \textsf{GCST}. \textsf{CST\_FULLY} is three orders of magnitude slower on this operation, taking over a millisecond. Interestingly, the larger representations, including those where the tree topology is represented using 2.79 bits per node (\textsf{CST\_SADA[\_RLCSA]}), are only marginally faster than \textsf{BT-CST}, whereas the indexes \textsf{CST\_SCT3[\_RLCSA]} are a bit slower than \textsf{CST\_SADA[\_RLCSA]} because they do not store an explicit tree topology. Note that these operations, in BT-CT, make use of the operations \fwdsearch\ and \bwdsearch, thereby showing that they are fast although we cannot prove worst-case upper bounds on their time.

Operation lca, which on \textsf{BT-CST} involves essentially the primitive \minexcess, is costlier, taking around 10 $\mu$sec in almost all the indexes, including ours. This includes again those where the tree topology is represented using 2.79 bits per node (\textsf{CST\_SADA[\_RLCSA]}). Thus, although we cannot prove upper bounds on the time of \minexcess, it is in practice is as fast as on perfectly balanced structures, where it can be proved to be logarithmic-time. The variants \textsf{CST\_SCT3[\_RLCSA]} also require an operation very similar to \minexcess, so they perform almost like \textsf{CST\_SADA[\_RLCSA]}. For this operation, \textsf{CST\_FULLY} is equally fast, owing to the fact that operation lca is a basic primitive in this representation. Only \textsf{GCST} is several times slower than \textsf{BT-CST}, taking several tens of $\mu$sec.

Operation suffix-link involves \minexcess\ and several other operations on the topology, but also the operation $\Psi$ on the corresponding CSA. Since the latter is relatively fast, \textsf{BT-CST} also takes nearly 10 $\mu$sec, whereas the additional operations on the topology drive \textsf{GCST} over 100 $\mu$sec, and \textsf{CST\_FULLY} over the millisecond. This time the topology representations that are blind to repetitivess are several times faster than \textsf{BT-CST}, taking a few $\mu$sec, possibly because they take more advantage of the smaller ranges for \minexcess\ involved when choosing random nodes (most nodes have small ranges). The \textsf{CST\_SCT3[\_RLCSA]} variants also solve this operation with a fast and simple formula.

Finally, operation child is the most expensive one, requiring one application of string-depth and several of next-sibling and letter, thereby heavily relying on the CSA. \textsf{BT-CST-bin} and \textsf{CST\_SCT3[\_RLCSA]} binary search the children; the others scan them linearly. The indexes using a CSA that adapts to repetitiveness require nearly one millisecond on large alphabets, whereas those using a larger and faster CSA are up to 10 (\textsf{CST\_SCT3}) and 100 (\textsf{CST\_SADA}) times faster. Our \textsf{BT-CST-bin} variant is faster than the base \textsf{BT-CST} by 15\% on \texttt{einstein} and 18\% on \texttt{kernel}, and outperforms the RLCSA-based indexes.
On DNA, instead, most of the indexes take nearly 100 $\mu$sec, except for \textsf{CST\_SADA}, which is several times faster; \textsf{GCSA}, which is a few times slower; and \textsf{CST\_FULLY}, which stays in the millisecond.

\begin{figure}[t]
    \centering
    \includegraphics[scale=0.35]{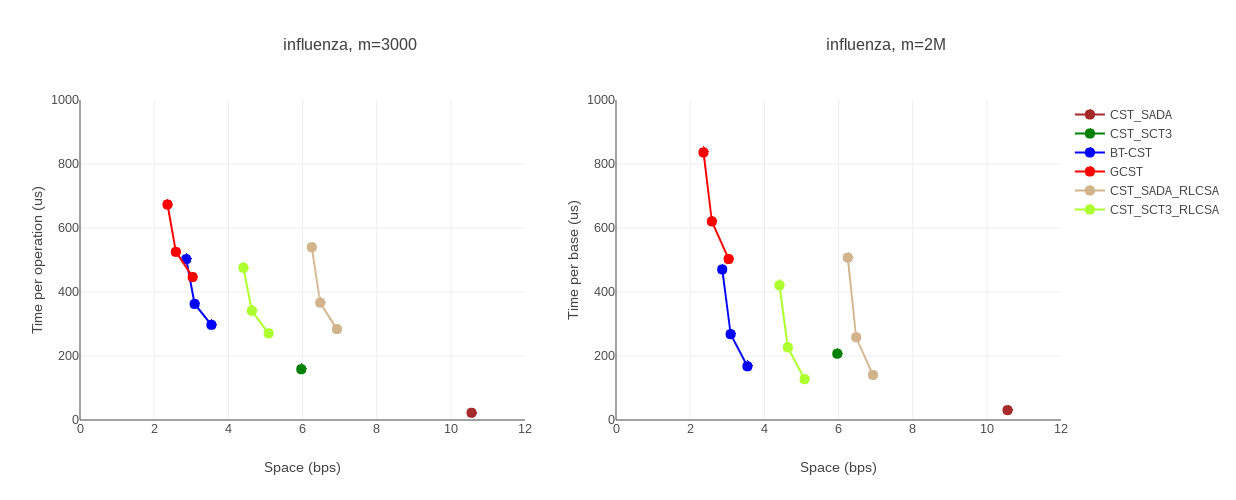}

    \vspace*{-4mm}
    \caption{Performance of CSTs when solving the maximal substrings problem. The y-axis is time in microseconds per base in the smaller sequence (of length $m$).}
    \label{maximal}
\end{figure}

Figure~\ref{maximal} shows the results for the maximal substrings problem. \textsf{BT-CST} sharply dominates an important part of the Pareto-curve, including the sweet point at 3.5 bps and 200-300 $\mu$sec per symbol. The other structures for repetitive collections take either much more time and slightly less space (\textsf{GCSA}, 1.5--2.5 times slower), or significantly more space and slightly less time (\textsf{CST\_SCT3}, 45\% more space and around 200 $\mu$sec). \textsf{CST\_SADA} is around 10 times faster, the same as its CSA when solving the dominant operation, child.

\section{Future Work}



Although we have shown that in practice they perform as well as their classical counterpart \cite{navarro2014fully}, an interesting open problem is whether the operations $\fwdsearch$, $\bwdsearch$, and $\minexcess$ can be supported in polylogarithmic time on Block Trees. This was possible on perfectly balanced trees~\cite{navarro2014fully} and even on balanced-grammar parse trees~\cite{navarro2016faster}, but the ability of Block Trees to refer to a prefix or a suffix of a block makes this more challenging. We note that the algorithm described by Belazzougui et al.~\cite{belazzougui2015queries} claiming logarithmic time for $\minexcess$ does not really solve the operation (as checked with coauthor T. Gagie).

\end{document}